\documentclass[12pt]{article}
\usepackage{epsf}
\usepackage{a4}
\usepackage[table]{xcolor}
\usepackage{cite}
\usepackage{avant}
\usepackage{amssymb}
\usepackage{multirow,booktabs}
\usepackage[all]{pstricks}
\usepackage{pst-plot,pst-math,pst-grad}
\usepackage{pifont}
\usepackage[fleqn]{amsmath}
\usepackage{caption}
\usepackage{subcaption}
\usepackage[mathscr]{eucal}
\usepackage[all]{xy}
\usepackage{listing}
\usepackage{pdfpages}
\usepackage{graphicx}
\usepackage[rightcaption]{sidecap}
\usepackage{mdframed}
\xyoption{arc}
\xyoption{curve}
\numberwithin{equation}{section}
\numberwithin{table}{section}
\newfont{\spfnt}{punk12}
\newcommand\email[4]{#1@#2.#3.#4}

\def\ber{\begin{eqnarray}}
\def\eer{ \end{eqnarray}}  
\def\be{\begin{equation}}
\def\ee{\end{equation}}
\def\la{\label}

 \def\n{\nonumber\\}

\def\bigone{\hbox{1\kern -.23em {\rm l}}}     
\def\ZZ{\hbox{\zfont Z\kern-.4emZ}}

\def\T{\Theta}

\catcode`@=11 \@addtoreset{equation}{section} \catcode`@=12

\begin{document}

\setlength \arraycolsep{2pt}
 
\title{Hyperscaling violating geometry with magnetic field and DC conductivity}
\author{
Subir Mukhopadhyay
\thanks{\email{subirkm}{gmail}{com}{}} 
and Chandrima Paul\thanks{\email{plchandrima}{gmail}{com}{}} \\
\small Department of Physics, Sikkim University, 6th Mile, Gangtok 737102
}
\date{}
\maketitle
\begin{abstract}
\small
\noindent
We consider dyonic black hole in hyperscaling violating Lifshitz theories arised in a four dimensional Einstein-Maxwell-dilaton system along with axion fields. Considering the linearised equation of relevant fluctuations in metric and gauge fields, we analytically compute thermoelectric conductivity of the dual theory using Dirichlet boundary condition and find agreement with conductivities obtained in near horizon analysis. We also study temperature dependence of the conductivities.
\end{abstract}
\thispagestyle{empty}
\clearpage


\section{Introduction}

Holographic techniques has been proved to be quite successful in analysing strongly coupled systems arised in condensed matter \cite{Hartnoll:2009sz,Herzog:2009xv,Horowitz:2010gk,Horowitz:2008bn,Hartnoll:2011fn,Sachdev:2011wg,Green:2013fqa,Hartnoll:2016apf}. In the original proposal\cite{Maldacena:1997re,Gubser:1998bc,Witten:1998qj}  it was for asymptotically anti de-Sitter spacetime and thus are amenable to theories at the boundary characterised by relativistic invariance at the boundary. Soon it transpires it can be generalised to other asymptotic spacetimes as well \cite{Kachru:2008yh,Taylor:2008tg,Ross:2009ar,Ross:2011gu,Mann:2011hg,Baggio:2011cp,Griffin:2011xs,Chemissany:2014xsa}. In particular, this has been extended to systems having anisotropic scaling symmetry along temporal and spatial direction. For such systems, asymptotically Lifshitz spacetimes turns out to be the pertinent set up on the gravity side. An essential motivation for these is to understand the novel behaviour of strongly correlated phases of matter, which cannot be explained using conventional theories, as it does not show quasiparticle description. Application of holographic methods for such phases are expected to provide new insights and deeper understanding about dynamics of these systems.

In this vein, a number of works have considered non-relativistic geometries which are asymptotically Lifshitz theories characterised by hyperscaling violation \cite{Cremonini:2016avj,Bhatnagar:2017twr,Cremonini:2018jrx,Ge:2016lyn,Chen:2017gsl}. A four dimensional Einstein-Maxwell-Axion-Dilaton theory gives rise to such geometries characterised by two parameters $z$ and $\theta$,  corresponding to Lifshitz scaling and the hyperscaling violation respectively. The Axion is chosen linear in space coordinates to introduce inhomogeneity in order to model the feature of underlying lattice structure \cite{Donos:2013eha,Andrade:2013gsa,Caldarelli:2016nni}. It involves two $U(1)$ gauge fields, one of which is required to introduce Lifshitz like behaviour, other playing the role of electromagnetic field. 

Electrically  charged black hole background in this theory has been considered and electrical DC conductivity was computed \cite{Cremonini:2016avj} using near horizon analysis \cite{Donos:2014cya}. In \cite{Bhatnagar:2017twr}, a magnetic field has been introduced in addition and thermoelectric conductivity was studied using near horizon analysis, once again. However, near horizon analysis \cite{Donos:2014cya}, though very useful, does not provide the conserved current in the boundary theory. In addition, it is not flexible to incorporate different boundary conditions of the fields in the bulk. Instead, it chooses one boundary condition out of multiple possibilities.

In view of these, a different approach has been proposed in \cite{Cremonini:2018jrx}. It considered linearised fluctuations around the electrically charged black hole and from analysis of asymptotic behaviour of the solutions they determine counterterms, obtain the physical observables in the dual theory and compute the thermoelectric conductivities. Unlike near horizon analysis, this approach is amenable to incoroporate different boundary conditions on the fields, such as Dirichlet and Neumann or a combination of them. 

In the present work, we have extended the approach of \cite{Cremonini:2018jrx} in presence of magnetic field. We consider a dyonic black hole background and from the analysis of linear fluctuations of necessary fields we have computed the full thermal conductivity matrix. This provides the dependence on magnetic field $B$ and in particular enable one to compute Hall angles. This analysis can accommodate different boundary coditions which may lead to different behaviour of thermal conductivities. In the present case, we have used Dirichlet boundary condition on spatial components of one of the gauge fields and find agreement of conductivities derived in approach of near horizon analysis \cite{Bhatnagar:2017twr}. We have discussed  temperature dependence of thermoelectric cunductivities and Hall angle in several scaling regimes.

This paper is structured as follows. In the next section we introduce the  asymptotically Lifshitz hyperscaling violating solution. In the section 3 we introduce the fluctuations in metric and gauge fields, consider their linearised equations of motion and obtain solution in low frequency limit.  In section 4 we compute the thermoelectric coefficients and discuss their temperature dependence. We conclude in section 5. Some of the materials related to the necessary canonical transformation of the fields has been discussed in the appendix.

\section{Hyperscaling violating Lifshitz Black Hole}

In the present section we will discuss the asymptotically Lifshitz hyperscaling violating solution, which we will use as the background. The electrically charged solution has been discussed in \cite{Cremonini:2016avj,Cremonini:2018jrx} and the dyonically charged solution has been mentioned in \cite{Ge:2016lyn,Bhatnagar:2017twr}. They appear as a classical solution of an Einstein-Maxwell-dilaton-axion system. We will consider two gauge fields coupled through a symmetric invertible matrix $\Sigma_{IJ}$, $I,J=1,2$ which is a function of the dilaton $\phi$, having positive eigenvalues. In addition, there are two axion fields, $\chi^a$, with $a$ running over $1,2$ required to violate the momentum conservation, which is necessary for computation of direct conductivity. The Axion term in the action has a dilaton dependent prefactor $Z(\phi)$.  

The four dimensional action is given by
\begin{equation}\label{action}
S = \int d^4x \sqrt{-g} [ R - \alpha (\partial\phi)^2 - \Sigma_{IJ} F_{\mu\nu}^I F^{J \mu\nu} - Z(\phi) (\partial\chi^a)^2 - V(\phi) ] + \frac{1}{2\kappa^2} \int\limits_{\mathcal \partial M} d^3x \sqrt{-\gamma} 2 K, 
\end{equation} 
where $\kappa^2$ in the second Gibbons-Hawking boundary term is given by  $8\pi G$. We consider two axion fields  and two guage fields with $I=1,2$.  $V(\phi)$ is the potential, which is functions of dilaton fields.

From the action (\ref{action}) we get the following equation of motion. The Einstein, Maxwell and dilaton, axion equations are
\begin{equation} \begin{split}\label{einstein}
&R_{\mu\nu} = \alpha \partial_\mu\phi \partial_\nu\phi + \frac{1}{2} V(\phi) g_{\mu\nu} + Z(\phi) \partial_\mu\chi^a \partial_\nu\chi^a + 2 \Sigma_{IJ}(\phi) (F^I_{\mu\lambda} F^{J\lambda}_\nu - \frac{1}{4} g_{\mu\nu} F^I_{\rho\sigma} F^{J\rho\sigma} ),\\&
\nabla^\mu (\Sigma_{IJ}(\phi) F_{\mu\nu}^J) = 0,\\&
\nabla^\mu(Z(\phi)\partial_\mu \chi^a) = 0 
\quad \text{and} \quad
2 \alpha \Box\phi - V^\prime(\phi) = \Sigma_{IJ}^\prime(\phi) F^I_{\rho\sigma} F^{J\rho\sigma},
\end{split}\end{equation}
respectively. 

In order to obtain asymptotically Lifshitz hyperscaling violating solution we choose the following ans$\ddot{\text{a}}$tz for the metric, axion and the gauge fields.
\begin{equation}\begin{split} \label{ansatz}
ds_B^2 &= \gamma_{\mu\nu} dx^\mu dx^\nu = dr^2 + e^{2A} ( - f(r) dt^2 + dx^a dx^a ),\\
\chi^a_B &= px^a,\quad\quad
 \phi_B = \phi_B(r), \quad\quad
A^I = a^I = a_t^I(r) dt + \frac{B^I}{4} \epsilon_{ab} x^a dx^b ,
\end{split}
\end{equation}
Where $\gamma_{ab}$ denotes background metric tensor. We have chosen a linear axion to break the translation invariance to incorporate momentum relaxation. The first gauge field is required to generate a Lifshitz like behaviour of the metric. while the second one gives rise to the dyonic charge of the solution. For the sake of generality, we have kept the constant magnetic field $F^I_{ab} = \frac{1}{2}B^I\epsilon_{ab}$ associated with both the gauge fields.   

Substituting the ansatz (\ref{ansatz}) in the second equation of (\ref{einstein}) implies the elctric charges $q_I = - f^{-1/2} e^A \Sigma_{IJ}  \partial_r a_t^J$ is  constant. The first and the last equation (\ref{einstein}), on substitution of the ansatz (\ref{ansatz}) reduces to the following equations:
\begin{equation}\begin{split}\label{eom-bg}
\frac{f^{\prime\prime}}{2 f} + 3 A^\prime \frac{f^\prime}{2 f} - \frac{f^{\prime 2}}{4f^2} = p^2 Z(\phi) e^{-2A} + 2  e^{-4A} ( \Sigma^{IJ}(\phi) q_I q_J + \frac{1}{4} \Sigma_{IJ}(\phi) B^I B^J ),\\  
A^{\prime\prime} + A^\prime (3 A^\prime + \frac{f^\prime }{2 f} ) + p^2 Z(\phi) e^{-2A} + \frac{1}{2} V + e^{-4A} ( \Sigma^{IJ} (\phi)q_I q_J + \frac{1}{4} \Sigma_{IJ}(\phi) B^I B^J ),\\
(6A^{\prime 2} + 4 A^\prime \frac{f^\prime}{2 f} ) = \alpha (\partial_r\phi)^2 - 2 p^2 Z(\phi) e^{-2A} - V - 2  e^{-4A} ( \Sigma^{IJ} (\phi)q_I q_J + \frac{1}{4} \Sigma_{IJ}(\phi) B^I B^J ),\\ 
2\alpha [ \partial_r^2 \phi + (3 A^\prime + \frac{f^\prime }{2 f} )\partial_r\phi ] - V^\prime(\phi) = 2  e^{-4A} ( \Sigma^{IJ\prime}(\phi)(\phi) q_I q_J + \frac{1}{4} \Sigma_{IJ}^\prime(\phi) B^I B^J ).
\end{split}\end{equation}
Given a form of $Z(\phi)$ and $\Sigma_{IJ}(\phi)$ one can solve these equations to find out the metric, the Maxwell field, the dilaton and the potential.

Like the electrically charged black hole, these equations do admit an exact dyonic black hole solution \cite{Bhatnagar:2017twr,Ge:2016lyn}, which depends on two parameters $z$ and $\theta$. We present the solution in radial coordinate $v$, which is particularly suited for asymptotic behavior. The metric in terms of this radial coordinate $v$ is given by 
\begin{equation}
ds^2 = v^{-\theta} [ - v^{2z} F(v) dt^2 + \frac{dv^2}{v^2 F(v)^2} + v^2 (dx^2 + dy^2)],
\end{equation}
where in our ansatz (\ref{ansatz}) we set $e^{2A} = v^{2-\theta}$ and  the blackening factor $F(v)$ is given by
\begin{equation}
F(v) = 1 + \frac{p^2}{(2-\theta)(z-2) v^{2z-\theta}} - \dfrac{m}{v^{2+z-\theta}} + \dfrac{8 q_2^2}{(2-\theta)(z-\theta) v^{2(z+1-\theta)}} + \dfrac{B^2 v^{2 z - 6}}{16 (4 + \theta -3 z)(2-z)}.\end{equation}
In terms of $v$ coordinate, the role of the blackening factor is played by $F(v)$.
This $v$ coordinate is related to $r$ through
\begin{equation}
dr = -sgn(\theta) v^{-\theta/2}{\mathcal F}^{-1/2}(v) \frac{dv}{v}.
\end{equation} 

Other fields and functions are given as follows: $\Sigma_{IJ}(\phi)$ and $Z(\phi)$ are
\begin{equation}
\Sigma_{11}(\phi) = \frac{1}{4} e^{[(\theta - 4)/\mu]\phi}
,\quad
\Sigma_{22}(\phi) = \frac{1}{4} e^{[(2z - 2 - \theta)/\mu]\phi}
, \quad
\Sigma_{12} =  0
, \quad
Z(\phi) = \frac{1}{2} e^{[\mu/(\theta -2)]\phi},
\end{equation}
where $\alpha = 1/2$ and $\mu$ is given by $2 \mu^2 \alpha = (2-\theta)(2z - 2 -\theta)$.
The dilaton, the axion and the gauge fields are given by
\begin{equation}
\phi = \mu \log v 
,\quad
\chi^a = p x^a
,\quad
a_t^1 = \frac{4 sgn(\theta) q_1 }{2+z-\theta} ( v^{2+z-\theta} -  v_h^{2+z-\theta} )
,\quad
a_t^2 = \frac{4 sgn(\theta) q_2 }{\theta - z} ( v^{\theta - z} -  v_h^{\theta - z} ).
\end{equation}
The charge $q_1$ and the potential $V(\phi)$ are
\begin{equation}
q_1^2 = (2+z-\theta)(z-1)/8
,\quad
V(\phi) = - (2+z-\theta)(1+z-\theta) e^{\theta \phi /\mu} - \frac{2z -2 -\theta}{4(z-2)} B^2 e^{(\theta + 2 z -6)(\phi/\mu)}.
\end{equation}

Unless otherwise mentioned we will keep our analysis general without commiting to specific solution. The reason is as follows. For electrically charged case, $B^I=0$ it can be shown that general solution with asymptotic behaviour exists. We expect a similar general solution with specific asymptotic behaviour in the case of dyonic black hole, as well. Therefore the present set up may be used to deal with general solutions. Though, while studying the coefficients of conductivities we will use the specific exact solution only.

\section{Fluctuation}

We will be interested in the thermoelectric coefficients, which are related to the correlation function of operators. In order to compute those we consider linear fluctuations in the metric and the gauge fields around its background solution.
\begin{equation}
\gamma_{ij} = \gamma_{Bij} + h_{ij},
\quad
A^I_i = A^I_{Bi} + a^I_i ,
\quad
\phi = \phi_B + \varphi ,
\quad
\chi^a = \chi_B^a + \tau^a,
\end{equation}
where $i,j$ takes values on $t$, $x$ and $y$.
Defining $S_i^j = \gamma^{jk}h_{ik}$, one can set  $S_t^t=S_x^x=S_y^y=S_x^y = 0$ and $\varphi = a_t^I =0$ consistently, leaving nonzero fluctuations to be $S_t^a$, $S_a^t$, $a_a^I$ and $\tau^a$. $S_a^t$ is related to $S_t^a$ and so we will not consider the former. In what follows, we will assume these fields depend on $t$ and $r$ only. With such dependence the linearised equations satisfied by these fluctuations for the background given in the ansatz (\ref{ansatz}) turn out to be as follows:
\begin{equation}\begin{split}
&[\partial_r^2 + (3 \partial_r A - \frac{\partial_r f}{2 f} )\partial_r - e^{-2A} ( 2 p^2 Z + e^{-2A} \Sigma_{IJ} B^I B^J) ]S_t^a = - 2 e^{-2A} [ pZ (\partial_t \tau^a )+ 2 \Sigma_{IJ} (\partial_r a_t^I ) (\partial_r a_a^J) \\ &
\hphantom{[\partial_r^2 + (3 \partial_r A - \frac{\partial_r f}{2 f} )\partial_r - e^{-2A} ( 2 p^2 Z + e^{-2A} \Sigma_{IJ} B^I B^J) ]S_t^a =}
+ e^{-2A}\Sigma_{IJ} (\partial_t a_b^I)\epsilon_{ab} B^J ],\\&
\partial_r\partial_t S_t^a + 2 e^{-2A} \Sigma_{IJ} (\partial_r a_t^I) B^J \epsilon_{ab} S^b_t = - 2 p f Z \partial_r\tau^a -4 e^{-2A} \Sigma_{IJ} \partial_r a_t^I \partial_t a_a^J - 2 f e^{-2A}\Sigma_{IJ}B^J \epsilon_{ab} \partial_r a_b^I \\&
\partial_r \{ \Sigma_{IJ} e^A f^{-1/2} [ (\partial_r a_t^J) S_t^a + f \partial_r a_a^J ]\} = f^{-1/2} e^{-A} \Sigma_{IJ} (\partial_t^2 a_a^J + \frac{1}{2} \epsilon_{ab} \partial_t S^b_t B^J,\\&
\partial_r^2\tau^a + (3 \partial_r A + \frac{\partial_r f}{2f} + \frac{\partial_r Z}{Z} ) \partial_r\tau^a - \frac{e^{-2A}}{f} \partial_r^2 \tau^a = - f^{-1} e^{-2A} p \partial_t S_t^a,
\end{split}\end{equation}
where we have not included equations for $S_a^t$, which follows from the above set of equations.
Considering the time dependence of the various functions is given by $e^{i\omega t}$, the above set of equations reduce to the following
\begin{equation}\begin{split}\label{eom1}
&[\partial_r^2 + (3 \partial_r A - \frac{\partial_r f}{2 f} )\partial_r - e^{-2A} ( 2 p^2 Z + e^{-2A} \Sigma_{IJ} B^I B^J) ]S_t^a 
= - 2 e^{-2A} [ -i\omega pZ \tau^a  + 2 \Sigma_{IJ} (\partial_r a_t^I ) (\partial_r a_a^J) \\&
\hphantom{[\partial_r^2 + (3 \partial_r A - \frac{\partial f}{2 f} )\partial_r - e^{-2A} ( 2 p^2 Z + e^{-2A} \Sigma_{IJ} B^I B^J) ]S_t^a = }
+ i\omega e^{-2A}\Sigma_{IJ} a_b^I \epsilon_{ab} B^J ],\\
&i\omega \partial_r S_t^a + 2 e^{-2A} \Sigma_{IJ} (\partial_r a_t^I) B^J \epsilon_{ab} S^b_t = - 2 p f Z \partial_r\tau^a -4 i \omega e^{-2A} \Sigma_{IJ} \partial_r a_t^I a_a^J - 2 f e^{-2A}\Sigma_{IJ}B^J \epsilon_{ab} \partial_r a_b^I, \\
&\partial_r \{ \Sigma_{IJ} e^A f^{-1/2} [ (\partial_r a_t^J) S_t^a + f \partial_r a_a^J ]\} = f^{-1/2} e^{-A} \Sigma_{IJ} (- \omega^2 a_a^J + \frac{i\omega}{2} \epsilon_{ab} S^b_t B^J),\\
&\partial_r[ e^{3A} f^{1/2} Z \partial_r \tau^a ] = - i\omega p Z e^A f^{-1/2} (S_t^a - \frac{i\omega}{p} \tau^a).
\end{split}\end{equation}
Following \cite{Cremonini:2018jrx} we introduce new field 
\begin{equation}
\Theta^a = S_t^a - \frac{i\omega}{p} \tau^a.
\end{equation} 
The boundary operator associated with $\Theta^a$ plays the role of energy operator in the boundary theory.
Introducing $\Omega = \omega^2 - 2 p^2 f Z$ we write down the equations in terms of this new field $\Theta^a$. Some of the terms, however, we have written in terms of  $S_t^a$, which can be expressed in terms of $\Theta^a$ and $\tau^a$. 
\begin{equation}\begin{split}\label{eom2}
&\partial_r [ 2p^2 f Z \Omega^{-1} (- f^{-1/2} e^{3A} \partial_r\Theta^a + 4 q_I a_a^I) - 2i\omega\Omega^{-1}B^I\epsilon_{ab} (q_I S_t^b - f^{1/2}e^A\Sigma_{IJ}\partial_ra_b^J)]\\& - e^Af^{-1/2} (2p^2Z + e^{-2A}\Sigma_{IJ} B^IB^J )\Theta^a = \frac{i\omega}{p} e^{-A} f^{-1/2} \Sigma_{IJ}B^IB^J \tau^a - 2i\omega e^{-A} f^{-1/2}\epsilon_{ab}\Sigma_{IJ}B^J a_b^I,\\
&f^{-1/2}e^A \partial_r (f^{1/2} e^A \Sigma_{IJ} \partial_r a_a^J - 2p^2fZ\Omega^{-1} q_I\Theta^a) - \frac{2 p^2 \omega^2}{\Omega^2} f^{-1/2}e^A \partial_r(fZ) q_I\Theta^a  + \omega^2 f^{-1} (\Sigma_{IJ} \\&- 4 \Omega^{-1} e^{-2A} f q_I q_J ) a_a^J + 2i\omega \Omega^{-1} e^{-2A}\epsilon_{ab} q_I B^J (q_J S_t^b - e^A f^{1/2}\Sigma_{JK} \partial_r a_b^K ) - \frac{i\omega}{2} f^{-1}\Sigma_{IJ}\epsilon_{ab}S_t^b B^J = 0,\\
&\partial_rS_t^a + 4 e^{-2A} \Sigma_{IJ}(\partial_r a_t^I)a_a^J = \frac{2i}{\omega} e^{-2A} \Sigma_{IJ} B^J \epsilon_{ab} (\partial_r a_t^J S_t^b + f \partial_r a_b^J) + \frac{2ipfZ}{\omega} \partial_r\tau^a,\\
&\partial_r [ e^{3A}f^{1/2}Z\partial_r\tau^a ] = -i\omega pZ e^A f^{-1/2} \Theta^a.
\end{split}\end{equation}

In order to obtain near horizon limit, we will use another radial coordinate $u$, which is related to $r$ through $du = - f(r)^{1/2} e^{-A(r)} dr$. In terms of $u$ the metric becomes
\begin{equation}
ds^2  = e^{2A(u)} ( - f(u) dt^2 + \frac{du^2}{f(u)} + dx^a dx^a ).
\end{equation}
The derivative in $u$ is related to that in $r$ through 
\begin{equation}
\partial_r = - \sqrt{f} e^{-A} \partial_u, \quad \partial_u = - f^{-1/2} e^A \partial_r
\end{equation} 
$u$ is related to $v$ through the relation $du = sgn(\theta) v^{z-3} dv$ where $z$ and $\theta$ are parameters determining behavior of the metric. 
The horizon of the black hole solution is given by $u=u_h$, where $f(u_h)=0$ and at the near horizon limit $f(r) \equiv 4\pi T \rho + {\mathcal O}(\rho^2)$, where $\rho = u_h - u$. $A$, $Z$ and $\Sigma_{IJ}$ approaches constant values at the near horizon limit. 
The near horizon limit of the four equations can be arranged in the following manner. 
\begin{equation}\begin{split}\label{horizon}
&\frac{2p^2Z}{\omega^2} [f\partial_u (f\partial_u (e^{2A}\Theta^a))] + 2p^2Z e^{2A} \Theta^a - \frac{2i}{\omega} \epsilon_{ab}\Sigma_{IJ} B^I [ f\partial_u(f\partial_u a_b^J) + \omega^2 a_b^J] + \frac{8p^2 Z}{\omega^2} q_I f\partial_u(f a_a^I) \\
&- \frac{2i}{\omega} q_I B^I \epsilon_{ab} \partial_u S_t^b + \Sigma_{IJ} B^IB^J S_t^a = 0,\\
&\Sigma_{IJ} [f\partial_u(f\partial_u a_a^J) + \omega^2 a_a^J] 
- \frac{2p^2Z q_I}{\omega^2} f^2 \partial_u\Theta^a 
- 4 e^{-2A} q_Iq_J f a_a^J 
 + \frac{2i}{\omega} e^{-2A} q_I B^J \epsilon_{ab} \Sigma_{JK} f^2 \partial_u a_b^K   \\
&+\frac{2i}{\omega} e^{-2A}\epsilon_{ab} q_I q_JB^J f S_t^b 
-\frac{i\omega}{2} \Sigma_{IJ}\epsilon_{ab} S_t^b B^J = 0,
\\
&\partial_u S_t^a - \frac{2i}{\omega} e^{-2A} \epsilon_{ab} q_IB^I S_t^b + 4 e^{-2A} q_I a_a^I - \frac{2i}{\omega} e^{-2A}\epsilon_{ab}\Sigma_{IJ}B^J f \partial_u a_b^J - \frac{2ip}{\omega} f \partial_u\tau^a = 0,
\\
&f\partial_u(f Z \partial_u e^{2A}\tau^a) = -i\omega p Z e^{2A}\Theta^a.
\end{split}\end{equation}
Considering the terms contributing in leading order of $\rho$ we obtain
\begin{equation}\begin{split}\label{horizon}
&\frac{2p^2Z}{\omega^2} [f\partial_u (f\partial_u (e^{2A}\Theta^a))] + 2p^2Z e^{2A} \Theta^a - \frac{2i}{\omega} \epsilon_{ab}\Sigma_{IJ} B^I [ f\partial_u(f\partial_u a_b^J) + \omega^2 a_b^J] +  \Sigma_{IJ} B^IB^J S_t^a = 0,\\
&\Sigma_{IJ} [f\partial_u(f\partial_u a_a^J) + \omega^2 a_a^J] 
-\frac{i\omega}{2} \Sigma_{IJ}\epsilon_{ab} S_t^b B^J = 0.
\end{split}\end{equation}
Introducing 
\begin{equation}
\eta_a^I = a_a^I + \frac{1}{2p} e^{2A} B^I \epsilon_{ab}\tau^b,
\end{equation}
and choosing the in-falling behaviour, we obtain the following near horizon behaviour
\begin{equation}\label{nh}
e^{2A}\Theta^a \sim \rho^{\frac{-i\omega}{4\pi T}} 
,\quad\quad
\eta_a^I  \sim \rho^{\frac{-i\omega}{4\pi T}}.
\end{equation}
We will use the above near horizon behaviour to determine the relations among the constants that appear in the solutions of the various fields.

In order to study direct conductivity, we require the solution of the fields $\Theta^a$, $a_a^I$ and $\tau^a$. However, the differential equations are quite involved and since we will be interested in the direct conductivity which depends on the behaviour of the fields at low frequency limit we will  expand the fields in powers of frequency and from there we will determine the low frequency behaviour of the fields.  So we consider the following expansions
\begin{equation}\label{expansion}
\Theta^a = \Theta^{a(0)} + \omega \Theta^{a(1)} + \omega^2 \Theta^{a(2)} + ...,
a_a^I = a_a^{I(0)} + \omega a_a^{I(1)} + \omega^2 a_a^{(2)} + ...,
\tau^a = \tau^{a(0)} + \omega \tau^{a(1)} + \omega^2 \tau^{a(2)} + ... .
\end{equation}
We will substitute these expansions in the equations and will determine the fields at different orders of frequency in an iterative manner. 

First we will consider the equations at the order of zero frequency. Substituting the expansions of (\ref{expansion}) in (\ref{eom2}) we obtain from the second equation in (\ref{eom2})
\begin{equation}\label{a0-eq}
\partial_r ( f^{1/2}e^A \Sigma_{IJ} \partial_r a_a^{J(0)} - q_I \Theta^{a(0)}) = 0,
\end{equation}
which suggests it is convenient to define a new function
\begin{equation}\label{c-def}
C_I^a =  f^{1/2} e^A \Sigma_{IJ} \partial_r a_a^J - q_I \Theta^a.
\end{equation} 
Then (\ref{a0-eq}) implies $C_I^{a(0)}$ is a constant.
From the first equation in (\ref{eom2})) we get
\begin{equation}\label{theta0-eq}
\partial_r [ e^{3A} f^{3/2} \partial_r(f^{-1} \Theta^{a(0)}) + 4 a_t^I C_I^{a(0)} ] = 0,
\end{equation}
where we have used the equation of background fields (\ref{eom-bg}).
From the third equation of (\ref{eom2}) one obtains for axion
\begin{equation}\label{tau0-eq}
\partial_u\tau^{a(0)} = \epsilon_{ab} \frac{C_I^{b(0)} B^I}{e^{2A}Z} f^{-1}.
\end{equation}

From (\ref{a0-eq}) and (\ref{theta0-eq}) we write the solutions in terms of integrals
\begin{equation}\begin{split} \label{soln-0}
\Theta^{a(0)} &= f \Theta_1^a + f \Theta_2^a \int \frac{du}{e^{2A} f^2} - 4 f C_I^{a(0)} \int \frac{a_t^I du}{e^{2A} f^2},\\
a_a^{I(0)} &= a^{I(0)}_{a0} - C_I^{a(0)} \int \frac{\Sigma^{IJ}}{f} du - q_J \Theta_1^a \int \Sigma^{IJ} du - q_J\Theta_2^a \int \Sigma^{IJ} \int \frac{du}{e^{2A} f^2}
\\& - 4 q_J C_K^{a(0)} \int du \Sigma^{IJ}\int \frac{a_t^K du}{e^{2A} f^2},\\
\tau^{a(0)} &= \tau^{a(0)}_0 + \epsilon_{ab} C_I^{b(0)} B^I \int \frac{du}{e^{2A} f Z} ,
\end{split}
\end{equation}
where $\Theta^a_1$, $\Theta^a_2$, $a^{I(0)}_{a0}$ and $\tau^{a(0)}_0$ are constants of integration.

At the near horizon limit, $A$, $Z$ and $\Sigma_{IJ}$ are approaching constant value $A(h)$, $Z(h)$ and $\Sigma_{IJ}(h)$. Behaviour of $f(u)$ near $u\rightarrow u_h$ is $f \sim 4\pi T \rho$ and $a^I_t \sim {\mathcal O}(\rho)$, which leads to
\begin{equation}\begin{split} \label{nh-0}
\Theta^{a(0)} = (4\pi T \rho)\Theta_1 + \frac{\Theta_2^a}{4\pi T e^{2A(h)}} - \frac{4C_I^{a(0)} \partial_u a_t^I}{4\pi T e^{2A(h)}} \rho \log\rho,\\
a_a^{I(0)} = a_{a0}^{I(0)} + (\frac{q_J \Theta_2^a}{4\pi T e^{2A(h)}} + C_J^{a(0)}) \frac{\Sigma^{IJ}(h)}{4\pi T} \log\rho + q_J \Theta_1^a \Sigma^{IJ}(h) \rho,\\
\tau^{a(0)} = \tau^{a(0)}_0 - \epsilon_{ab} \frac{C_I^{b(0)} B^I}{4\pi T e^{2A(h)} Z(h) } \log\rho.
\end{split}\end{equation}
The equations at the zeroeth order of frequency are very much similar to that obtained in absence of magnetic field\cite{Cremonini:2018jrx} as in the equations $B^I$ appears at the first order of $\omega$.  

Next we will consider the equations at first order of frequency. As we have already mentioned, we will use a recursive procedure to determine the solutions at different orders of $\omega$, by using solutions obtained in the lower orders. Substituting the (\ref{expansion})in the second equation in (\ref{eom2})) we get
\begin{equation}\begin{split}\label{c1-eq}
&f^{-1/2} e^A \partial_r (f^{1/2} e^A \Sigma_{IJ}\partial_r a_a^{J(1)} - q_I \Theta^{a(1)}) - \frac{2i}{2p^2fZ} e^{-2A}\epsilon_{ab} q_IB^J (q_J\Theta^{b(0)} - e^A f^{1/2} \Sigma_{JK}\partial_r a_b^{K(0)})\\ 
&- \frac{i}{2} f^{-1} \Sigma_{IJ} \epsilon_{ab} \Theta^{b(0)} B^J = 0,
\end{split}\end{equation}
which leads to
\begin{equation}\begin{split}\label{c1-eq-2}
\partial_u C_I^{a(1)} = -if^{-1}\epsilon_{ab}B^J (- \frac{q_I C_J^{b(0)}}{p^2 Z e^{2A}} + \Sigma_{IJ}\frac{\Theta^{b(0)}}{2}).
\end{split}\end{equation}
By integrating (\ref{c1-eq-2}) we can write $C_I^{a(1)} $ in terms of the zeroeth order terms.
Similarly, $\Theta^{a(1)}$ and $\tau^{a(1)}$ satisfy
\begin{equation}\begin{split}\label{theta1-eq}
&\partial_u[e^{2A} f^2 \partial_u (f^{-1}\Theta^{a(1)})] - 4 C_I^{a(1)} \partial_ua_t^I + 2i \epsilon_{ab} \Sigma_{IJ}B^J a_b^{I(0)} \\&
+ \frac{i}{p^2} \epsilon_{ab} B^I f C_I^{a(0)} \partial_u(\frac{1}{fZ}) - \frac{i}{p} \Sigma_{IJ} B^IB^J \tau^{a(0)} = 0,\\&
\partial_u[e^{2A} f Z \partial_u \tau^{a(1)}] = i p Z e^{2A}f^{-1} \Theta^{a(0)},
\end{split}\end{equation}
while $a_a^{I(1)}$ can be obtained from 
\begin{equation}\label{a1-eq}
\partial_u a_a^{I(1)} = - f^{-1} \Sigma^{IJ} C_J^{a(1)} - q_I f^{-1} \Theta^{a(1)}.
\end{equation}
Like $C_I^{a(1)} $, all these equations can be integrated to obtain expressions at first order in terms of the zeroeth order fields. 

The near horizon behaviour of the fields at first order can be obtained by integrating the above equations after substituting the near horizon behaviour of $f$, $A$, $Z$ and $\Sigma$ and using the expressions obtained for the zeroeth order fields. For $C_I^{a(1)}$ we obtain,
\begin{equation}\begin{split}
C_I^{a(1)} = &C_{I0}^{a(1)} + \frac{i\epsilon_{ab}B^J}{e^{2A(h)} 4\pi T} [ \left( -\frac{q_I C_J^{b(0)}}{p^2 Z(h)} + \frac{\Theta_2^b\Sigma_{IJ}(h)}{8\pi T} \right) \log\rho + \frac{1}{2}\Theta_1^a \Sigma_{IJ}(h)e^{2A(h)} 4\pi T \rho 
\\& + 
 \frac{2 \Sigma_{IJ} C_K^{a(0)} \partial_ua_t^K}{4\pi T} (\rho \log\rho - \rho)] + ...,
\end{split}\end{equation}
where $C_{I0}^{a(1)}$ is an integration constant.

Using this expression a similar near horizon expression can be obtained for $\Theta^{a(1)}$ from (\ref{theta1-eq}) as follows
\begin{equation}\begin{split}\label{theta1-sol}
\Theta^{a(1)} = \frac{\Theta^a_3}{e^{2A(h)} 4\pi T} + \frac{i}{p^2} \frac{\epsilon_{ab} B^I C_I^{a(0)}}{Z(h)} \log\rho  + \Theta_4^a 4\pi T \rho + ...,
\end{split}\end{equation}
where $\Theta_3^a$ and $\Theta_4^a$ are new integration constants.
The fluctuation in gauge field at first order, $a_a^{I(1)}$ at the near horizon limit follows from (\ref{a1-eq}) and is given by
\begin{equation}\label{a1-sol}
a_a^{I(1)} =  a_{a0}^{I(1)} + \frac{\Sigma^{IJ}(h)}{4\pi T} \left[ \frac{q_J \Theta_3^a}{e^{2A(h)} 4\pi T} + C_{J0}^{a(1)} \right] \log\rho + ....,
\end{equation}
where we have introduced the constant term of integration as $a_{a0}^{I(1)}$. Finally the $\tau^a$ at first order turns out to be
\begin{equation}\begin{split}
\tau^{a(1)} = \tau^{a(1)}_0 - \frac{e^{-2A(h)}}{4\pi T p Z(h)} \left[ \epsilon_{ab}B^K C_{K0}^{b(1)} - \frac{i}{2} (e^{2A} 4\pi T \Theta_1^a + 4 q_ia_{a0}^{I(0)}) + \frac{i}{p} q_I B^I \epsilon_{ab} \tau^{b(0)}_0 \right] \log\rho + ....,
\end{split}\end{equation}

The constants of integration introduced at different orders can be determined by comparing with the near horizon behaviour with the full fledged expressions of the various fluctuations, obtained in (\ref{nh}). For that we need to consider the equations to the second order in $\omega$. 

At the second order of $\omega$ we obtain the following equation for $C_I^{a(2)}$
\begin{equation}\begin{split}\label{C2-eq}
\partial_u C_I^{a(2)} = &\frac{q_I e^{-2A}}{2p^2fZ} [ (e^{2A} \partial_u\Theta^{a(0)} + 4 q_J a_a^{J(0)} )+ 2i \epsilon_{ab} (C_J^{b(1)} - \frac{i}{p} q_J\tau^{b(0)}) B^J ] \\&- \frac{\Sigma_{IJ}}{f} [ a_a^{J(0)} - \frac{i}{2} (\Theta^{b(1)} + \frac{i}{p}\tau^{b(0)})B^J]
\end{split}\end{equation}
On the other hand for $\Theta^{a(2)}$ we get
\begin{equation}\begin{split}\label{theta2-eq}
\partial_u[e^{2A} f^2 \partial_u (f^{-1}\Theta^{a(2)})] =  &4 C_I^{a(2)} \partial_u a_t^I - f \partial_u [ \frac{1}{2p^2fZ}[ (e^{2A} \partial_u\Theta^{a(0)} + 4 q_J a_a^{J(0)} )
\\+ 2i \epsilon_{ab} (C_J^{b(1)} -& \frac{i}{p} q_J \tau^{b(0)}) B^J ] + \frac{i}{p} \Sigma_{IJ}B^IB^J \tau^{a(1)} - 2i \epsilon_{ab} a_b^{I(1)}.
\end{split}\end{equation}
$a_a^{I(2)}$ can be obtained as usual, from
\begin{equation}\label{a2-eq}
C_I^{a(2)} =  f^{1/2} e^A \Sigma_{IJ} \partial_r a_a^{J(2)} - q_I \Theta^{a(2)}.
\end{equation} 

In order to compare to the boundary condition at horizon we need to find the leading order behaviour of the fields near the horizon. Substituting the expressions we have obtained for fields upto zeroeth order and first order on right hand side of (\ref{C2-eq}) one can easily find that the leading order terms of $C_I^{a(2)}$ near horizon are of the order of $\log\rho$ and $(\log\rho)^2$. In particular, it does not have any $1/\rho$ in its expression near the horizon. It follows from equation for $\Theta^{a(2)}$  that the leading order expression of $\Theta^{a(2)}$ is given by
\begin{equation}
f^{-1} \Theta^{a(2)} = \Theta_6 + \Theta_5 \int \frac{du}{e^{2A}f^2} + S \log\rho + ...,
\end{equation}
where $\Theta_5^a$ and $\Theta_6^a$ are constants of integration and $S$ is given by
\begin{equation}
S = \frac{1}{2p^2Z(h)} [ (-4\pi T e^{2A} \Theta_1^a + 4 q_I a_{a0}^I ) + 2i\epsilon_{ab} (C^{b(1)}_{I0} - \frac{i}{p} q_I \tau^{b(0)}_0 ) B^I ]+ \frac{\Theta_2^a}{(4\pi T)^2}.
\end{equation}

Collecting expressions of $\Theta^a$ at different orders of frequency together, we can write near horizon expression of $\Theta^a$ valid upto ${\mathcal O}(\omega^2)$ as
\begin{equation}\begin{split}
\Theta^a = &\frac{\Theta_2^a}{e^{2A(h)} 4\pi T} + 4\pi T \Theta_1^a \rho + (   \frac{1}{4\pi T e^{2A(h)} }
[ \frac{i\omega\epsilon_{ab}C_{I0}^b} {p^2Z(h)} + \frac{2 (    -\pi T e^{2A(h)}\Theta_1^a + q_I a_{a0}^{I0})}   {p^2 Z(h)} \omega^2\\& +\frac{\omega^2}{p^3 Z(h)} \epsilon_{ab}\tau^{b(0)}_0 q_I B^I ] + \omega^2 \frac{\Theta_2^a}{(4\pi T)^2} )\log\rho + ...
\end{split}\end{equation}
In this equation, following \cite{Cremonini:2018jrx} we have absorbed all the pertinent integration constants in $\Theta_1^a$, $\Theta_2^a$ and $C_{I0}^a$, without any loss of generality, by redefining $\Theta_2^a$,  $\Theta_1^a$ and $C_{I0}^a$. 
 Similarly the expression for the fluctuation in gauge field at near horizon limit is 
 \begin{equation}
 a_a^I = a_{a0}^I + \frac{\Sigma^{IJ}(h)}{4\pi T} (C_{J0}^a + \frac{q_J\Theta_2^a}{e^{2A(h)} 4\pi T} ) \log\rho + ...
 \end{equation}
 where we have absorbed all the constants of integration in
$ a_{a0}^I$. Fluctuation in the axion $\tau^a$ at near horizon turns out to be
\begin{equation}
\frac{\tau^a}{p} = \frac{\tau^{a}_0}{p} + \frac{1}{4\pi T e^{2A(h)} p^2 Z(h)} [ - \epsilon_{ab}  B^I C_{I0}^b + 2i\omega (-\pi T e^{2A}(h) \Theta_1^a + q_I a_{0a}^I ) + i\omega (q_IB^I) \epsilon_{ab} \frac{\tau^{b(0)}_0}{p} ] \log\rho + ... 
\end{equation}
where constants are absorbed in $\tau^{a}_0$.

Comparing with the near horizon behaviour of $\Theta^a$ and $\eta_a^I = a_a^I + \frac{1}{2} B^I \epsilon_{ab} \frac{\tau^b}{p}$ as given in (\ref{nh}), we obtain
\begin{equation}\begin{split} \label{relation-constants}
(\Sigma^{IJ}(h) + \frac{B^I B^J }{2 p^2 Z e^{2A(h)}})C_{J0}^a + \frac{\Sigma^{IJ}(h) q_J}{e^{2A(h)} 4\pi T} \Theta_2^a &=  i \omega \{ (a_{a0}^I + \frac{1}{2} B^I \epsilon_{ab} \frac{\tau^b_0}{p} ) 
- \frac{B^I}{e^{2A(h)} p^2 Z} [ \epsilon_{ab} 
\\&
(-\pi T e^{2A(h)} \Theta_1^b  + q_I a^I_{0b}) 
+ \frac{1}{2} (q_JB^J) \frac{\tau^a_0}{p} \}, \\
\Theta_2^a =  - \frac{4\pi T}{p^2Z(h)} [ \epsilon_{ab} B^I C_{I0}^b  - 2 i\omega (-\pi T e^{2A(h)} &\Theta_1^a + q_I a_{a0}^I ) -i\omega (q_JB^J) \epsilon_{ab} \frac{\tau^b_0}{p} ].
\end{split}\end{equation}

From the two equations above (\ref{relation-constants}) we can express $C_{I0}^a$ and $\Theta_2^a$ in terms of other constants $a^I_{a0}$, $\Theta_1^a$ and $\tau^a_0$ in the following manner,
\begin{equation}\begin{split}
C_{I0}^a &= i\omega (M_I^{~J})_{ab} \{ - [(\Sigma_{JK}(h) + \frac{ 2 q_Jq_K}{p^2Z(h) e^{2A(h)}})\delta_{bc} + \frac{\Sigma_{JN}(h)B^N q_K}{p^2Z(h) e^{2A(h)}} \epsilon_{bc} ] a^k_{c0} + \frac{2\pi T}{p^2 Z(h)} [ q_J \delta_{ab} 
\\&+ \frac{1}{2} \Sigma_{JN}(h)B^N \epsilon_{bc} ] \Theta_1^c - \frac{1}{2} [ (\Sigma_{JK}(h) + \frac{4 q_J q_K}{p^2 Z(h) e^{2A(h)}} )B^K \epsilon_{bc} - (q_M B^M) \frac{\Sigma_{JK}(h)B^K}{p^2Z(h) e^{2A(h)}} \delta_{bc} ] \frac{\tau^c_0}{p} \},
\\
\Theta_2^a &= -\frac{4\pi T}{p^2 Z(h)} \epsilon_{ab} B^I C_{I0}^b + i\omega \frac{4\pi T}{p^2Z(h)} [ 2 (-\pi T e^{2A(h)} \Theta_1^a + q_I a_{0a}^I ) + (q_IB^I)\epsilon_{ab} \frac{\tau_0^b}{p}],
\end{split}\end{equation}
upto leading order in $\omega$, where we have introduced the matrix $(M_I^J)_{ab}$ satisfying
\begin{equation}\label{defM}
[(\delta_I^J + \frac{\Sigma_{IN}(h) B^N B^J }{2p^2Z(h) e^{2A(h)}})\delta_{ab} - \frac{q_IB^J}{p^2Z(h) e^{2A(h)}} \epsilon_{ab} ] (M_J^{~K})_{bc} = \delta_I^K\delta_{ac}.
\end{equation}
In absence of magnetic field it reduces to $\delta_I^J\delta_{ab}$.

In order to identify the operators in the boundary theory, we require the asymptotic solution of $\Theta^a$, $a_a^I$ and $\tau^a$. It is sufficient to determine the asymptotic solution of the fields upto lowest order in frequency. From the linearised equations of motion of the fluctutations it is clear that magnetic field contributes at a higher order in frequency. Therefore, upto lowest order of frequency, expressions remain the same as those obtained in absence of magnetic field \cite{Cremonini:2018jrx}. To this end we introduce
\begin{equation}\begin{split}\label{asymptotic}
\Psi(v) &= sgn(\theta) \int \frac{ v^{\theta-3z-1} dv}{F(v)^2},\\
Y^1(v) &= \frac{4 sgn(\theta) q_1}{2+z-\theta}( - v_h^{2+z-\theta} \Psi(v) + sgn(\theta)\int dv v^{-2z+1}F^{-2}), \\
Y^2(v) &= \frac{4 sgn(\theta) q_2}{\theta - z}( - v_h^{\theta - z} \Psi(v) + sgn(\theta)\int dv v^{2\theta - 4z - 1}F^{-2}). \\
\end{split}\end{equation}
In terms of these functions we can write the asymptotic expansions of the solutions of the fields at small frequency 
\begin{equation}\begin{split}
\label{asymptotic}
\Theta^{a(0)} &= v^{2(z-1)} F(v) ( \Theta_1^a +  \Theta_2^a \Psi(v) + 4 C_I^a Y^I(v)), \\
a_a^I &= a_{a0}^I - \Theta_1^a a^I_t - sgn(\theta) \Theta_2 q_J \int dv \Sigma^{IJ} v^{z-3} \Psi(v)\\& - sgn(\theta) \int dv \Sigma^{IJ} v^{-z-1} ( F^{-1} \delta_K^J + 4q_J v^{2(z-1)} Y^K(v) ) C_K^a .
\end{split}
\end{equation}

From (\ref{asymptotic}) one can establish a relation between the parameters describing the asymptotic behaviour of the solutions and  operators in the boundary theory. This relation has been discussed elaborately in \cite{Cremonini:2018jrx} and we have included their discussion in the appendix. As explained there, a basis of symplectic variables that parametrize the asymptotic solutions can be identified from asymptotic behaviour of the generalised coordinates and momenta. To this end one considers the radial Hamiltonian formulation and express asymptotic solutions of the linear fluctuations of the fields $\Theta^{a(0)}$, $a_a^{I(0)}$ and $\tau^{a(0)}$ and their conjugates in terms of the modes $\Theta_1^a$, $ \Theta_2^a$, $a_{a0}^I$, $C_I^a$ and $\tau^a_0$. Then one makes a suitable  canonical transformation, that can be realised by adding appropriate counterterms, leading to holographic renormalisation of the action. From the asymptotic behaviour of these transformed canonical variables the operators can be identified in terms of the modes parametrizing the asymptotic solution. 
 
Choice of the boundary condition turns out to play critical role in this identification. As explained in \cite{Cremonini:2018jrx} adding an additional finite term in the renormalised on-shell action, the Dirichlet boundary condition can be imposed on the gauge field. In the case of electrically charged black hole as the background, it has been found that the expressions of the conductivities obtained using the near horizon method agrees with the Dirichlet boundary condition. In the present case, where we have magenteic field in addition, we are considering the Dirichlet boundary condition so as to compare the results already obtained using near horizon method. With the present set up generalising it to Neumann or mixed boundary condition is quite straightforward. 

In case of Dirichlet boundary condition, we are interested in energy operator ${\mathcal E}^a$ and current operator ${\mathcal J}_I^a$ as shown in \cite{Cremonini:2018jrx}. Their expressions in terms of different modes are given by (\ref{Ea}) and (\ref{JIa}) 
\begin{equation}\begin{split} \label{Ea1}
{\mathcal E}^a =  -\frac{1}{2\kappa^2} (\Theta_2^a + 4 \mu^I C_{I0}^a ),\quad
{\mathcal J}_I^a = -\frac{2}{\kappa^2} (C_{I0}^a - \frac{i\omega q_I}{p} \tau_0^a ),\quad
{\mathcal X}_a = - \frac{2i\omega}{p\kappa^2} q_I \alpha_a^I.
\end{split}\end{equation}
where $\alpha_a^I$ is obtained from the asymptotic behaviour for the renormalised variables as given in (\ref{alpha}).
From these expressions we can obtain the various correlation function, that leads to computation of the coefficients of thermoelectric conductivity.

\section{Thermoelectric DC conductivities}

In this section we obtain thermoelectric conductivities for the present model. In the last section we have derived $\Theta_2^a$ and $C_{I0}^a$ in terms of other constants in (\ref{relation-constants}). We substitute these expressions in the energy operator ${\mathcal E}^a$  given in (\ref{Ea1}), we get
\begin{equation}\begin{split}
&{\mathcal E}^a = - \frac{i\omega}{2\kappa^2} [ 
			\{ 
\frac{8 \pi T }{p^2 Z} q_K \delta_{ad} - (-\frac{4\pi T}{p^2 Z}\epsilon_{ab}B^I + 4 \mu^I \delta_{ab})(M_I^{~J})_{bc} [ (\Sigma_{JK} + \frac{4q_Jq_K}{p^2Z e^{2A} })\delta_{cd} + \frac{\Sigma_{JM}B^M q _k}{p^2Ze^{2A}}\epsilon_{cd}
			\} \alpha_{d0}^K
\\
			&+( 
\frac{8\pi T}{p^2 Z} (q_K\mu^K - \pi T e^{2A})\delta_{ad} + (  - \frac{4\pi T}{p^2Z} \epsilon_{ab}B^I + 4 \mu^I\delta_{ab}    ) (M_I^J)_{bc} 
	\{ 
	\frac{2\pi T }{2p^2 Z} (     q_J\delta_{cd} + \frac{1}{2} \Sigma_{JM}B^M \epsilon_{cd}     ) 
\\& 
- [     (  \Sigma_{JK} + \frac{2 q_Jq_K}{p^2Z e^{2A}   }  )\mu^K\delta_{cd} + \frac{\Sigma_{JM}B^M q_K\mu^K}{p^2Ze^{2A}} \epsilon_{cd}   ]
		\} 
		 )\Theta_1^d
 + \{ -\frac{1}{2} (- \frac{4\pi T}{p^2Z}\epsilon_{ab} B^I + 4\mu^I\delta_{ab})(M_I^{~J})_{bc}
 \\& [ (\Sigma_{JK} + \frac{4 q_Jq_K}{p^2Z e^{2A}}  )B^K\epsilon_{cd} - \frac{(q_KB^K) \Sigma_{JM}B^M}{p^2Ze^{2A}} \delta_{cd} ] + \frac{4\pi T}{p^2Z} (q_KB^K) \epsilon_{ad} \} \frac{\tau_0^d}{p}.
  \end{split}\end{equation}
  where we have used the asymptotic value of fluctuation in gauge field, $\alpha_a^I$ givn in (\ref{alpha}). In this section, to simplify the notation, unless otherwise mentioned $A$, $\Sigma_{IJ}$ and $Z$ represents their respective values at the near horizon limit. 
  
  Similarly, the current operator ${\mathcal J}_I^a$ turns out to be
  \begin{equation}\begin{split}
 &J_I^a = \frac{2i\omega}{\kappa^2} [ 
 (M_I^{~J})_{ab} [  (  \Sigma_{JK} + \frac{2 q_Jq_K}{p^2Z e^{2A}   }  )\delta_{bc} + \frac{\Sigma_{JM}B^M q_K}{p^2Ze^{2A}} \epsilon_{bc}  ]\alpha_{c0}^K 
 \\
& - (M_I^{~J})_{ab} \{ 
	\frac{2\pi T }{2p^2 Z} (     q_J\delta_{bc} + \frac{1}{2} \Sigma_{JM}B^M \epsilon_{bc}     ) 
\\& 
- [     (  \Sigma_{JK} + \frac{2 q_Jq_K}{p^2Z e^{2A}   }  )\mu^K\delta_{bc} + \frac{\Sigma_{JM}B^M q_K\mu^K}{p^2Ze^{2A}} \epsilon_{bc}   ]
		\} \Theta_1^c
		\\&
		+\{\frac{1}{2} (M_I^{~J})_{ab}[(  \Sigma_{JK} + \frac{2 q_Jq_K}{p^2Z e^{2A}   }  )B^K\epsilon_{bc} - (q_NB^N) \frac{\Sigma_{JK} B^K}{p^2Ze^{2A}}\delta_{bc}]  + q_I \delta_{ac} \}\frac{\tau_0^c}{p},
		\\
 &{\mathcal X}_a = - \frac{2i\omega}{p\kappa^2} q_I\alpha^I_a,
 \end{split}\end{equation}
 where the matrix $(M_I^J)_{ab}$ is given by (\ref{defM}).
 
 From the above expressions one can obtain the following two-point functions
 \begin{equation}\begin{split} 
 \langle {\mathcal J}_I^a(-\omega) {\mathcal J}_J^b(\omega) \rangle &= \frac{2i\omega}{\kappa^2} (M_J^{~K})_{bc} [(\Sigma_{KI} + \frac{2q_Kq_I}{p^2Ze^{2A}}) \delta_{ca} + \frac{\Sigma_{KM}B^M q_I}{p^Ze^{2A}} \epsilon_{ca} ],
 \\
 \langle {\mathcal E}^a(-\omega) {\mathcal J}_I^b(\omega) \rangle &= -\frac{2i\omega}{\kappa^2}(M_I^J)_{bc}
 \{
 \frac{2\pi T}{p^2 Z} (q_J \delta_{ca} + \frac{1}{2} \Sigma_{JK} B^K \epsilon_{ca} ) 
 \\&
 -  [    (  \Sigma_{JK} + \frac{2 q_Jq_K}{p^2Z e^{2A}   }  )\delta_{bc} + \frac{\Sigma_{JM}B^M q_K}{p^2Ze^{2A}} \epsilon_{bc}   ]\mu^K
\}, \\
 \langle {\mathcal J}_I^a(-\omega) {\mathcal E}^b(\omega) \rangle &=  \frac{2i\omega}{\kappa^2} [ ( -\frac{2 \pi T }{p^2 Z}q_I \delta_{ba} - ( - \frac{\pi T}{p^2 Z} B^J \epsilon_{bc} + \mu^J \delta_{bc} )
 (M_J^{~K})_{cd} \\
 &[ (\Sigma_{KI} + \frac{2q_K q_I}{p^2Ze^{2A}})\delta_{da} + \frac{\Sigma_{KM}B^M q_I}{p^2Z e^{2A}} \epsilon_{da} ] 
 \}, \\
\langle {\mathcal E}^a(-\omega) {\mathcal E}^b(\omega) \rangle &= \frac{2i\omega}{\kappa^2} [\frac{2\pi T}{p^2 Z} (q_K\mu^K - \pi T e^{2A} ) \delta_{ba} + ( - \frac{\pi T}{p^2 Z e^{2A} } \epsilon^{bc} B^I  + 4 \mu^I \delta_{bc})
\\& (M_I^J)_{cd} 
 [ (\frac{2\pi T }{p^2 Z} q_J - (\Sigma_{JK} + \frac{2 q_J q_K}{p^2 Z e^{2A}})\mu^K ]\delta_{da} + \Sigma_{JM}B^M ( \frac{\pi T}{p^2 Z } - \frac{q_K\mu^K}{p^2Ze^{2A}})\epsilon_{da} ],
  \\
\langle {\mathcal X}^a(-\omega) {\mathcal J}_I^b(\omega) \rangle &= \frac{2i\omega}{\kappa^2} [ \frac{1}{2} (M_I^J)_{bc} [ (\Sigma_{JK} + \frac{4 q_J q_K }{p^2 Z e^{2A}})B^K \epsilon_{ca}  - (q_JB^J) \frac{\Sigma_{JK} B^K}{p^2 Z e^{2A}}\delta^{ca} ] + q_I \delta_{ba},
\\
\langle {\mathcal J}_I^a(-\omega) {\mathcal X}^b(\omega)  \rangle &= - \frac{2i\omega}{p \kappa^2} q_I \delta^{ab},
\end{split}\end{equation}
with rest of the two point functions vanishing.

Next following \cite{Cremonini:2018jrx} we introduce the heat current
\begin{equation}
 {\mathcal Q}_D^a = {\mathcal E}^a - \mu^I {\mathcal J}_I^a.
\end{equation}
The two point function for heat current and electric currents are given by
\begin{equation}\begin{split}
 \langle {\mathcal Q}_D^a(-\omega) {\mathcal Q}_D^b(\omega) \rangle &= \frac{2i\omega}{\kappa^2} 2 ( \frac{\pi T }{p^2Z})^2 \{ p^2 Z e^{2A}\delta^{ab} + B^M \epsilon_{ac}(M_M^{~J})_{cd}[q_J \delta_{da} + \frac{1}{2} \Sigma_{JN}B^N \epsilon_{da} ] \},\\
  \langle {\mathcal Q}_D^a(-\omega) {\mathcal J}_I^b(\omega) \rangle &= - \frac{2i\omega}{\kappa^2} \frac{2 \pi T}{p^2 Z} [(M_I^{~J})_{bc} [ q_J \delta_{ca} + \frac{1}{2} \Sigma_{JK}B^K \epsilon_{ca}],\\
\langle {\mathcal J}_I^a(-\omega) {\mathcal Q}_D^b(\omega)  \rangle &= -\frac{2i\omega}{\kappa^2} \{  \frac{2\pi T}{p^2 Z} q_I \delta_{ba} + \frac{\pi T }{p^2 Z} \epsilon_{bc} B^J (M_J^{~K})^{cd} [ (\Sigma_{KI} + \frac{2q_Kq_I}{p^2Ze^{2A}})\delta_{da} + \frac{\Sigma_{KM} B^M q_J }{p^2Z E^2A} \epsilon_{da} ] \}\\
 \langle {\mathcal J}_I^a(-\omega) {\mathcal J}_J^b(\omega) \rangle &= \frac{2i\omega}{\kappa^2} (M_J^{~K})_{bc} [(\Sigma_{KI} + \frac{2q_Kq_I}{p^2Ze^{2A}}) \delta_{ca} + \frac{\Sigma_{KM}B^M q_I}{p^Ze^{2A}} \epsilon_{ca} ],
\end{split}\end{equation}

We obtain the thermoelectric conductivities from the above two point functions as follows.
\begin{equation}
{\mathcal \sigma}_D^{DC} = \left(\begin{array}{cc} T \bar{\mathbb K}^{ab}&T \bar{\mathcal\alpha}_I^{ab}\\T {\mathcal\alpha}_I^{ab}&{\mathcal\sigma}_{IJ}^{ab}\end{array}\right) 
= \left(\begin{array}{cc} \langle {\mathcal Q}_D^a(-\omega) {\mathcal Q}_D^b(\omega) \rangle&\langle {\mathcal Q}_D^a(-\omega) {\mathcal J}_I^b(\omega) \rangle \\
\langle {\mathcal J}_I^a(-\omega) {\mathcal Q}_D^b(\omega)  \rangle&\langle {\mathcal J}_I^a(-\omega) {\mathcal Q}_D^b(\omega)  \rangle \end{array}\right).
\end{equation}
In order to obtain the following expressions for the components of the conductivity matrix in a compact form we have introduced the following parameters 
\begin{equation}
r_I = \frac{1}{2} \Sigma_{IJ}B^J, \quad\quad b^I = \frac{B^I}{p^2Z e^{2A}}.
\end{equation}
In terms of these parameters the matrix $(M_I^{~J})_{ab}$ is given from (\ref{defM})
\begin{equation}
(M_I^{~J})_{ab} = \delta_I^J\delta_{ab} - \frac{[(1+r.b)r_I + (q.b)q_I ] \delta_{ab} - [(1+r.b)q_I - (q.b) r_I ]\epsilon_{ab} }{(1+r.b)^2 + (q.b)^2}b^J .
\end{equation} 
where we have used $(r.b) = r_Ib^I$, $(q.b)=q_Ib^I$ and $\triangle = (1+r.b)^2 + (q.b)^2$. With these expressions, components of conductivity matrix becomes
\begin{equation}\begin{split}
\bar{\mathbb K}^{ab} &=  \frac{\pi s T}{\kappa^2 p^2 Z } \frac{[(1+r.b) \delta_{ba} + (q.b) \epsilon_{ba}]}{\triangle},
\\
\bar{\alpha}_I^{ab} &= \alpha_I^{ab} = - \frac{4}{sT} \bar{\mathbb K}^{bc} (q_I\delta_{ca} + r_I \epsilon_{ca})],
\\
{\mathcal \sigma}_{IJ}^{ab} &= \frac{2}{\kappa^2} \Sigma_{JI}\delta^{ba} + \frac{16}{s^2 T} \bar{\mathbb K}^{bc} (q_J \delta_{cd} + r_J \epsilon_{cd} )(q_I \delta_{da} + r_I \epsilon_{da} ),
\end{split}\end{equation}
 where we have used $4 \pi e^{2A} = s$. All the components of the conductivity matrix reduce to the expressions of the same given in \cite{Cremonini:2018jrx} for setting $B^I=0$ It may be observed that both the $U(1)$ gauge fields are on the same footing and that we have got $\bar{\alpha}_I^{ab} = \alpha_I^{ab}$. 

We have obtained the thermoelectric conductivities for the general case and in this form the symmetry between and electric and magnetic fields is also becomes apparent. We can apply this general result to the case of dyonic black hole discussed in section 2. Substituting values of the various quantities in the above expressions we obtain the following forms for conductivities. For the solution we get $\triangle = (p^2 + \frac{B^2}{4} v^{4z-6-\theta} )^2 + (2 q_2 B v^{2z-4})^2$ and using that we get,
\begin{equation}\begin{split}
\bar{\mathbb K}^{ab} &= \frac{8\pi^2 T}{\kappa^2 p^2} v_h^{2(z-\theta)} \frac{(p^2 + \frac{B^2}{4} v_h^{4z-6-\theta} ) \delta_{ba} + 2 q_2 B v_h^{2z-4} \epsilon_{ba}}{\triangle},\\
{\mathbb \alpha}_1^{ab} &= -\frac{8\pi}{\kappa^2} v_h^{2z-\theta-2} \frac{(p^2 + \frac{B^2}{4} v_h^{4z-6-\theta}) q_1 \delta_{ba} + 2 q_1 q_2 B v_h^{2z-4} \epsilon_{ba}}{\triangle},\\
{\mathbb \alpha}_2^{ab} &= -\frac{8\pi}{\kappa^2} v_h^{2z-\theta-2} \frac{p^2 q_2 \delta_{ba} + [(p^2 + \frac{B^2}{4} v_h^{4z-6-\theta} ) \frac{B}{8} v_h^{2z-2-\theta} + 2 q_2^2 B v_h^{2z-4}] \epsilon_{ba}}{\triangle},\\
{\mathcal \sigma}_{11}^{ab} &= \frac{1}{2\kappa^2} v_h^{\theta-4} \delta_{ba} + \frac{8}{\kappa^2} q_1^2 v_h^{2z-4} \frac{(p^2 + \frac{B^2}{4} v_h^{4z -6-\theta})\delta_{ba} + 2 q_2 B v_h^{2z-4} \epsilon_{ba}}{\triangle},\\
{\mathcal \sigma}_{12}^{ab} &= \frac{8}{\kappa^2} q_1 \frac{q_2p^2 \delta_{ba} + [2 q_2^2 B v_h^{2z-4} + \frac{B}{8} v_h^{2z-2-\theta} (p^2 + \frac{B^2}{4} v_h^{4z-6-\theta})]\epsilon_{ba}}{\triangle},\\
{\mathcal \sigma}_{22}^{ab} &= \frac{p^2}{2\kappa^2} v_h^{6z-8-2\theta}\frac{  \frac{B^2}{4} + v_h^{6-4z+\theta} (p^2 + 16 q_2^2 v_h^{\theta - 2})}{\triangle} \delta_{ba} + \frac{q_2 B }{\kappa^2} v_h^{8z-12-2\theta} \frac{\frac{B^2}{4} + (2p^2 + 16 q_2^2 v_h^{\theta - 2})v_h^{-4z+6+\theta}}{\triangle}\epsilon_{ba},
\end{split}\end{equation}
  
Hall angle can be obtained from the above conductivities by taking the ratio of coefficients of $\epsilon_{ab}$ and $\delta_{ab}$ in the expression of ${\mathcal\sigma}$. We get
  \begin{equation}
  \Theta_H = \frac{2q_2 B}{p^2} v_h^{2z-4}[ \frac{\frac{B^2}{4} + v_h^{-4z+6+\theta} (2 p^2 + 16 q_2^2 v_h^{\theta -2})}{\frac{B^2}{4} + v_h^{-4z+6+\theta} ( p^2 + 16 q_2^2 v_h^{\theta -2})}].
  \end{equation}
  As explained in \cite{Blake:2014yla} since the factor in the square bracket lies between 1 and 2 Hall coefiicient can be approximated as
  \begin{equation}
  \Theta_H = \frac{2q_2 B}{p^2} v_h^{2z-4}{ p^2}.
  \end{equation}
  these expressions,after setting $\theta=1-z$, agree with the results obtained in \cite{Bhatnagar:2017twr} using the near horizon method.
  
With the explicit expressions of various components of thermoelectric matrix we can study temperature dependence. For the analytic black hole solution the temperature is given by $T= - \frac{sgn(\theta)}{4\pi} v_h^{z+1} F'(v_h)$ which for the case of dyonic solution reduces to
  \begin{equation}
  T = - \frac{sgn(\theta)}{4\pi} [ (z+2-\theta) v_h^z - \frac{8q_2^2}{2-\theta} v_h^{2\theta-z-2} - \frac{p^2}{2-\theta} v_h^{\theta - z} - \frac{B^2}{4(2-z)} v_h^{3z-6}.
  \end{equation}
The expression of temperature is quite involved and it is difficult to obtan an analytic expression of the conductivities in terms of the temperature. 
Nevertheless, choosing appropriate limits of the quantities we can identify regimes, where one can discuss scaling behaviour of the coefficients with the temperature.

We begin with $\theta < 0$, where the first term is positive while rest of the terms are negative in the expression of temperature. To identify a regime of large temperature, following\cite{Cremonini:2018jrx} we consider $q_2^2 v_h^{2\theta-z-2} << v_h^z$, $p^2 v_h^{\theta-z} << v_h^z$ and $B^2 v_h^{3(z-2)} << v_h^z$. In this regime one can identify $T\equiv \frac{8q_1^2}{4\pi(z-1)} v_h^z$. The behaviour of thermoelectric conductivity matrix will depend on the relative strengths of the different terms in the temperature. We have considered the following three regions of parameters. Apart from that one can also obtain the cases, where two terms are comparable, but there it is difficult to identify the scaling behaviour of the conductivities.

We begin with the range of parameters where momentum dissipaion is strong compared to charge and magnetic field, which is given by, $B^2 v_h^{3(z-2)} , q_2^2 v_h^{2\theta-z-2} <<  p^2 v_h^{\theta-z} << v_h^z$. In this limit we obtain 
\begin{equation}\begin{split}\label{p-dominate}
{\mathbb K}^{ab} &\sim \frac{8\pi^2 T}{\kappa^2 p^4} [T^{\frac{2(z-\theta)}{z}} \delta_{ba} + 2 q_2 B T^{\frac{4z-2\theta-4}{z}} \epsilon_{ab}],\\
{\mathcal\sigma}_{11}^{ab} &\sim \frac{8q_1^2}{\kappa^2 p^2} [T^{\frac{2z -4}{z}} \delta_{ba} + \frac{2q_2 B}{ p^2} T^{\frac{4z-8}{z}}\epsilon_{ba}],\\
{\mathcal\sigma}_{12}^{ab} &\sim \frac{ q_1}{\kappa^2 p^2} [8 q_2 \delta_{ba} +   B T^{\frac{4z-6-\theta}{z}} \epsilon_{ba}]\\
{\mathcal\sigma}_{22}^{ab} &\sim \frac{1}{2\kappa^2} [ T^{\frac{2z-2-\theta}{z}}\delta_{ba} + \frac{4q_2B}{p^2} T^{\frac{4z-6-\theta}{z}} \epsilon_{ba}],\\
{\mathcal \alpha}_1^{ab} &\sim -\frac{8\pi q_1}{\kappa^2 p^2 } [ T^{\frac{2z - \theta-2}{z}} \delta_{ba} + \frac{2q_2 B}{p^2} T^{\frac{2z-4}{z}} \epsilon_{ba} ],\\
{\mathcal \alpha}_2^{ab} &\sim -\frac{8\pi}{\kappa^2 p^2 } [ q_2 T^{\frac{2z - \theta-2}{z}} \delta_{ba} + \frac{B}{8} T^{\frac{4z - 2\theta - 4}{z}} \epsilon_{ba} ].
\end{split}\end{equation}
The Hall angle is $\theta_H \sim T^{\frac{2z-4}{z}}$. Since $\theta < 0$ we cannot get linear resistivity for $\sigma_{22}^{xx}$ in this regime. Choosing $z=1$ we get $\theta_H \sim 1/T^2$ and $\sigma_{22}^{xx} \sim T^{-\theta}$ showing a positive power of $T$ for conductivity. Instead if we choose, $B^2 v_h^{3(z-2)} <<  p^2 v_h^{\theta-z} << q_2^2 v_h^{2\theta-z-2}<< v_h^z, p^2 >> 2q_2B v_h^{2z-4}$ all the coefficients will remain the same except $\sigma_{22}$. It becomes
\begin{equation}
\sigma_{22}^{ab} = \frac{8 q_2^2}{\kappa^2 p^2} [ T^\frac{2z-4}{z} \delta_{ba}  + \frac{2 q_2 B}{p^2} T^\frac{4z-8}{z}\epsilon_{ba}].
\end{equation}
In this regime, $\sigma_{22}^{xx}$ and Hall angle have similar temperature dependence. So for $z=1$ both scale as $\sim T^{-2}$. Choosing $z=4/3$ one gets $\sigma_{22}^{xx} \sim T^{-1}$  implying linear resistivity. However, Hall angle also becomes $\theta_H \sim T^{-1}$.
 
 Another scaling regime, that one may consider corresponds to the range where the charge is strong compared to momentum dissipation and magnetic field. That is given by $B^2 v_h^{3(z-2)} , p^2 v_h^{\theta-z} << q_2^2 v_h^{2\theta-z-2}<< v_h^z$ and leads to the following conductivities:
  \begin{equation}\begin{split}\label{q-dominate}
{\mathbb K}^{ab} &\sim \frac{8\pi^2 T}{\kappa^2 (2q_2B)} [\frac{1}{2 q_2 B} T^{\frac{2(4-\theta -z)}{z}} \delta_{ba} + \frac{1}{p^2} T^{\frac{2(2-\theta)}{z}} \epsilon_{ab}],\quad \text{for}\quad B^2 v_h^{3(z-2)}  <<  p^2 v_h^{\theta-z},
\\
{\mathcal \alpha}_1^{ab} &\sim -\frac{8\pi q_1}{\kappa^2 } [ \frac{p^2}{4q_2^2B^2} T^{\frac{6-2z - \theta}{z}} \delta_{ba} + \frac{1}{2q_2 B} T^{\frac{2-\theta}{z}} \epsilon_{ba} ] \quad \text{for}\quad B^2 v_h^{3(z-2)}  <<  p^2 v_h^{\theta-z}, \\
{\mathcal \alpha}_1^{ab} &\sim -\frac{8\pi q_1}{\kappa^2 } [ \frac{4}{B^2} T^{\frac{2(z - \theta)}{z}} \delta_{ba} + \frac{1}{2q_2 B} T^{\frac{2-\theta}{z}} \epsilon_{ba} ]\quad \text{for}\quad  p^2 v_h^{\theta-z} << B^2 v_h^{3(z-2)}  ,
\\
{\mathcal \alpha}_2^{ab} &\sim -\frac{8\pi}{\kappa^2 } [ \frac{p^2}{4q_2 B^2} T^{\frac{6-2z - \theta}{z}} \delta_{ba} + \frac{1}{2B} T^{\frac{2 - \theta }{z}} \epsilon_{ba} ],
\\
{\mathcal\sigma}_{11}^{ab} &\sim \frac{q_1^2}{2\kappa^2 } [\frac{4 p^2}{q_2^2 B^2}T^{\frac{4-2z}{z}} \delta_{ba} + \frac{16}{2q_2 B} \epsilon_{ba}],\quad \text{for}\quad  p^2 v_h^\theta >> (q_2 B v_h^{z+\theta-4})^2 ,\\
 &\sim \frac{q_1^2}{2 \kappa^2 } [T^{\frac{\theta - 4}{z}} \delta_{ba} + \frac{16}{2q_2 B} \epsilon_{ba}],\quad \text{for}\quad  p^2 v_h^\theta << (q_2 B v_h^{z+\theta-4})^2  ,
 \\
{\mathcal\sigma}_{12}^{ab} &\sim \frac{8 q_1 q_2}{\kappa^2} [\frac{p^2}{4 q_2^2 B^2} T^{\frac{8-4z}{z}} \delta_{ba} +   \frac{1}{2q_2B} T^\frac{4-2z}{z} \epsilon_{ba}],
\\
{\mathcal\sigma}_{22}^{ab} &\sim \frac{1}{2\kappa^2} [\frac{4p^2}{B^2} T^{\frac{4-2z)}{z}}\delta_{ba} + \frac{8 q_2}{B} \epsilon_{ba}].
\end{split}\end{equation}
In this regime, $\sigma_{22}^{xx}$ and Hall angle have opposite temperature dependence. Choosing $z=1$ one gets temperature dependence to be $T^2$ and $T^{-2}$ respectively. For $z=2$, however both will be independent of temperature. Similarly one can consider the regime where magnetic field will be stronger compared to the momentum dissipation and charge. In that regime, $\sigma_{22}^{xx} \sim T^{\frac{(4-2z)}{z}}$ with Hall angle having opposite temperature dependence, once again. 

For small temperature, one can identify the following regions of parameters. \\
$B^2 v_h^{3(z-2)} , q_2^2 v_h^{2\theta-z-2} <<  p^2 v_h^{\theta-z} \lesssim v_h^z$, $B^2 v_h^{3(z-2)} , p^2 v_h^{\theta-z} << q_2^2 v_h^{2\theta-z-2} \lesssim v_h^z$ and \\ $p^2 v_h^{\theta-z} , q_2^2 v_h^{2\theta-z-2}<< B^2 v_h^{3(z-2)}  \lesssim v_h^z$ .   However, obtaining an analytical expression for temperature for this region is difficult. The dependence on $v_h$ can be obtained from above by replacing $T$ by $v_h^z$ in (\ref{p-dominate}) and (\ref{q-dominate}) a respectively in the three regimes. 
 
 For $\theta > 0$ first term is negative and so large temperature may corresponds to the regimes depending on whether $p^2 v_h^{\theta-z}$, $q_2^2 v_h^{2\theta-z-2}$ or $B^2 v_h^{3(z-2)}$ dominates. In these regimes, temperature can be approximated by $T\equiv \frac{p^2}{4\pi(2-\theta)} v_h^{\theta - z}$, $T\equiv \frac{8 q_2^2}{4\pi(2-\theta)} v_h^{2\theta - z - 2}$ or $T\equiv \frac{ B_2^2}{16\pi(2-z)} v_h^{3z - 6}$, respectively. The scalings of conductivity matrix for various regimes will be as follows:
 
 For the parameter region corresponding to strong momentum dissipation, $B^2 v_h^{3(z-2)}$ , $q_2^2 v_h^{2\theta-z-2} <<  p^2 v_h^{\theta-z}$ we get
\begin{equation}\begin{split}\label{p-dominate1}
{\mathbb K}^{ab} &\sim \frac{8\pi^2 T}{\kappa^2 p^4} [\left(\frac{T}{p^2}\right)^{\frac{2(z-\theta)}{\theta -z}} \delta_{ba} + 2 \frac{q_2 B}{p^2} \left(\frac{T}{p^2}\right)^{\frac{4z-2\theta-4}{\theta-z}} \epsilon_{ab}],
\\
{\mathcal \alpha}_1^{ab} &\sim -\frac{8\pi q_1}{\kappa^2 p^2 } [ \left(\frac{T}{p^2}\right)^{\frac{2z - \theta-2}{\theta-z}} \delta_{ba} + \frac{2q_2 B}{p^2} \left(\frac{T}{p^2}\right)^{\frac{2z-4}{\theta-z}} \epsilon_{ba} ],
\\
{\mathcal \alpha}_2^{ab} &\sim -\frac{8\pi}{\kappa^2 p^2 } [ q_2 \left(\frac{T}{p^2}\right)^{\frac{2z - \theta-2}{\theta-z}} \delta_{ba} + \frac{B}{8} \left(\frac{T}{p^2}\right)^{\frac{4z - 2\theta - 4}{\theta-z}} \epsilon_{ba} ],
\\
{\mathcal\sigma}_{11}^{ab} &\sim \frac{8q_1^2}{\kappa^2 p^2} [\left(\frac{T}{p^2}\right)^{\frac{2z -4}{\theta-z}} \delta_{ba} + \frac{2q_2 B}{ p^2} \left(\frac{T}{p^2}\right)^{\frac{4z-8}{\theta - z}}\epsilon_{ba}],
\\
{\mathcal\sigma}_{12}^{ab} &\sim \frac{ q_1}{\kappa^2 p^2} [8 q_2 \delta_{ba} +   B \left(\frac{T}{p^2}\right)^{\frac{2z-2-\theta}{\theta - z}} \epsilon_{ba}],
\\
{\mathcal\sigma}_{22}^{ab} &\sim \frac{1}{2\kappa^2} [ \left(\frac{T}{p^2}\right)^{\frac{2z-2-\theta}{\theta-z}}\delta_{ba} + \frac{4q_2B}{p^2} \left(\frac{T}{p^2}\right)^{\frac{4z-6-\theta}{\theta-z}} \epsilon_{ba}.
\end{split}\end{equation}
For $z\rightarrow 2$ $\sigma_{22}^{xx} \sim T^{-1}$, but Hall angle becomes independent of temperature.

For the regime, where charge is strong compared to other two factors, given by $B^2 v_h^{3(z-2)}$ , $p^2 v_h^{\theta-z} << q_2^2 v_h^{2\theta-z-2}$, conductivities turn out to be
  \begin{equation}\begin{split}\label{q-dominate}\nonumber
{\mathbb K}^{ab} &\sim \frac{8\pi^2 T}{\kappa^2 p^2} [\frac{p^2}{4 q_2^2 B^2} \left(\frac{T}{q_2^2}\right)^{\frac{8-2z-2\theta}{2\theta-z-2}} \delta_{ba} + \frac{1}{2 q_2 B} \left(\frac{T}{q_2^2}\right)^{\frac{2(2-\theta)}{2\theta-z-2}} \epsilon_{ab}],\quad \text{for}\quad B^2 v_h^{3(z-2)}  <<  p^2 v_h^{\theta-z},\\
 &\sim \frac{8\pi^2 T}{\kappa^2 p^2} [\frac{1}{16 q_2^2} \left(\frac{T}{q_2^2}\right)^{\frac{2z + 2 - 3\theta}{2\theta-z-2}} \delta_{ba} + \frac{1}{2 q_2 B} \left(\frac{T}{q_2^2}\right)^{\frac{2(2-\theta)}{2\theta-z-2}} \epsilon_{ab}],\quad \text{for}\quad  p^2 v_h^{\theta-z} << B^2 v_h^{3(z-2)}  ,
 \\
 {\mathcal \alpha}_1^{ab} &\sim -\frac{8\pi q_1}{\kappa^2 } [ \frac{p^2}{4q_2^2B^2} \left(\frac{T}{q_2^2}\right)^{\frac{6-2z - \theta}{2\theta-z-2}} \delta_{ba} + \frac{1}{2q_2 B} \left(\frac{T}{q_2^2}\right)^{\frac{2-\theta}{2\theta-z-2}} \epsilon_{ba} ] \quad \text{for}\quad B^2 v_h^{3(z-2)}  <<  p^2 v_h^{\theta-z}, 
 \\
{\mathcal \alpha}_1^{ab} &\sim -\frac{8\pi q_1}{\kappa^2 } [ \frac{1}{16 q_2^2} \left(\frac{T}{q_2^2}\right)^{\frac{2(z - \theta)}{2\theta-z-2}} \delta_{ba} + \frac{1}{2q_2 B} \left(\frac{T}{q_2^2}\right)^{\frac{2-\theta}{2\theta-z-2}} \epsilon_{ba} ]\quad \text{for}\quad  p^2 v_h^{\theta-z} << B^2 v_h^{3(z-2)}  ,
\\
{\mathcal \alpha}_2^{ab} &\sim -\frac{8\pi}{\kappa^2 } [ \frac{p^2}{4q_2 B^2} \left(\frac{T}{q_2^2}\right)^{\frac{6-2z - \theta}{2\theta-z-2}} \delta_{ba} + \frac{1}{2B} \left(\frac{T}{q_2^2}\right)^{\frac{2 - \theta }{2\theta-z-2}} \epsilon_{ba} ].
\\
{\mathcal\sigma}_{11}^{ab} &\sim \frac{8q_1^2}{\kappa^2 } [\frac{p^2}{4q_2^2 B^2}\left(\frac{T}{q_2^2}\right)^{\frac{4-2z}{2\theta-z-2}} \delta_{ba} + \frac{1}{2q_2 B} \epsilon_{ba}],\quad \text{for}\quad p^2 v_h^\theta >> (2 q_2B v_h^{z+\theta-4})^2,
\\
&\sim \frac{1}{2 \kappa^2 } [\left(\frac{T}{q_2^2}\right)^{\frac{\theta - 4}{2\theta-z-2}} \delta_{ba} + \frac{1}{2q_2 B} \epsilon_{ba}],\quad \text{for}\quad p^2 v_h^\theta << (2 q_2B v_h^{z+\theta-4})^2,
\\
 &\sim \frac{8q_1^2}{\kappa^2 } [\frac{1}{16 q_2^2}\left(\frac{T}{q_2^2}\right)^{\frac{2z-2-\theta}{2\theta-z-2}} \delta_{ba} + \frac{1}{2q_2 B} \epsilon_{ba}],\quad \text{for}\quad  p^2 v_h^{\theta-z} << B^2 v_h^{3(z-2)}  ,
 \end{split}\end{equation}
 \begin{equation}\begin{split}\label{q-dominate1}
{\mathcal\sigma}_{12}^{ab} &\sim \frac{8 q_1}{\kappa^2} [\frac{p^2}{4 q_2 B^2} \left(\frac{T}{q_2^2}\right)^{\frac{8-4z}{2\theta-z-2}} \delta_{ba} +   \frac{1}{2B} \left(\frac{T}{q_2^2}\right)^{\frac{4-2z}{2\theta-z-2}}\epsilon_{ba}],
\\
{\mathcal\sigma}_{22}^{ab} &\sim \frac{1}{2\kappa^2} [\frac{4p^2}{B^2} \left(\frac{T}{q_2^2}\right)^{\frac{4-2z}{2\theta-z-2}}\delta_{ba} + \frac{8q_2}{B} \epsilon_{ba}].
\end{split}\end{equation}
As observed from above, $\sigma_{22}^{xx}$ and Hall angle has opposite temperature dependence. for For $z=1$ $\sigma_{22}^{xx} \sim T^{-1}$, but Hall angle becomes independent of time. Small temperature limit can be chosen in a similar way as in the case of $\theta < 0$. The bahaviour will be similar to those obtained in the case of $\theta < 0$.
 
 We have seen the behaviour of the various thermoelectric coefficients depends on competing contributions from different terms. For high temperature limits we  have discussed several regimes where the scaling with temperature can be identified. For small temperature, however, the dependence is quite involved and it is difficult to identify the behaviour with specific powers of temperature. In general, a numerical procedure can be used for obtaining temperature dependence.
 
 \section{Conclusion}
 
We have used holographic techniques to analyze thermoelectric properties of systems dual to hyperscaling violating Lifshitz geometry. Considering a dyonically charged black hole as the background we have turned on necessary fluctuations in metrics and gauge fields. Solving the equations of motion of the fluctuations and imposing in-falling boundary condition at the horizon we have obtained the thermoelectric coefficients from the asymptotic behaviour of fluctuations in low frequency limit. Compared to the near horizon method, this method \cite{Cremonini:2018jrx} has the advantage that it enables one to identify the boundary operators explicitly and is amenable to accommodate different boundary conditions. 

We have discussed the temperature dependence of various thermoelectric coefficients. Because of the background solution is too involved, we can analytically discuss only a few specific regimes. In one of the regimes, $z=4/3$ leads to linear resistivity but Hall angle goes as $1/T$, though for $z=1$ it shows  $1/T^2$ behaviour.  Here we have explicitly consider the dyonic background. It may be interesting to obtain the result in the case of electrically charged background, by using mixed boundary condition on the gauge field. A natural extension of the present work is to explore AC conductivity using numerical techniques and study temperature dependence for intermediate frequencies. Another direction is to consider turning on mass for the bulk gauge field \cite{Gouteraux:2012yr},
which gives rise to additional exponents.  The present method may also be applied to explore properties of the other models towards obtaining agreement with experimental observations.

 \appendix
 \section{Appendix}

 In order to determine the thermoelectric DC conductivities in this method we need to identify the operators in the boundary theory with the parameters describing the asymptotic behaviour of the solutions. These has been elaborated in \cite{Cremonini:2018jrx} and in this appendix we include a brief review for convenience. First we will consider a new set of coordinates parametrizing ``dual frame'', where radial coordinate is ${\bar r}$, which is related to the Einstein frame radial coordinate $r$ through the relation $d\bar{r} = - sgn(\theta) e^{\frac{\theta}{2\mu}\phi } dr$. The advantage of this dual coordinate is it allows both positive and negative values of $\theta$ and the UV boundary lies at ${\bar r} \rightarrow \infty$.
 
 In order to identify the operators living in the boundary theory and the fields in the bulk theory one considers \cite{Cremonini:2018jrx,Chemissany:2014xsa}  the symplectic set of variables consisting generalised coordinates and its canonically conjugate momenta in the bulk Hamiltonian radial formalism. This enables one to identify the natural basis of symplectic variables that parametrize the space of asymptotic solutions.  

The metric in the Einstein or the dual frame can be decomposed in the following manner.
$ds^2 = dr^2 + \gamma_{ij} dx^i dx^j$, where $x^i = {t, x^a}$. In the Hamiltonian formalism the metric and the gauge field can be decomposed as
\begin{equation}
ds^2 = (N^2 + N_iN^i) dr^2 + 2 N_i dr dx^i + \gamma_{ij} dx^i dx^j, \quad A^I_\mu dx^\mu = A^I_r dr + A^I_i dx^i,
\end{equation}
where $N$ and $N_i$ are the lapse and shift function and $\gamma_{ij}$ is the induced metric on radial slices at fixed values of $r$. Similarly $A_r$ and $A_i$ are transverse and longitudinal components of the gauge fields to the radial slices. 
We also write down the extrinsic curvature, which can be expressed as
\begin{equation}
K_{ij} = \frac{1}{2N} (\partial_r\gamma_{ij} - D_iN_j - D_jN_i),
\end{equation}
where $D_i$ is the covariant derivative with respect to the metric $\gamma_{ij}$. 
We will use barred quantities for dual frame and unbarred one for Einstein frame.

The lagrangian in the dual frame, as obtained in \cite{Cremonini:2018jrx} is given by
\begin{equation}\begin{split}
L_\xi &= \frac{1}{2\kappa^2} \int d^3 x \sqrt{-\bar\gamma} \bar{N} [ (1+ \frac{4\xi^2}{\alpha_\xi})\bar{K}^2 - \bar{K}^{ij}\bar{K}_{ij} - \frac{\alpha_\xi}{\bar{N}^2} ( \partial_r\phi - \bar{N}^i\partial_i\phi - \frac{2\xi}{\alpha_\xi}\bar{N}\bar{K} )^2
 \\
&- \frac{2}{\bar{N}^2} \Sigma^\xi_{IJ}(\phi) (F^I_{ri} - \bar{N}^k F^I_{ki} ) (F^{Ji}_r - \bar{N}^l F^{Ji}_l ) - \frac{1}{\bar{N}^2} Z_\xi(\phi) (\partial_r\chi^a - \bar{N}^i\partial_i\chi^a)^2
 \\
&+ R[\bar\gamma] - \alpha_\xi \partial_i\phi\bar{\partial}^i\phi - \Sigma^\xi_{IJ} F^I_{ij} F^{Jij} - Z_\xi \partial_i\chi^a \bar{\partial}^i\chi^a -V_\xi - 2 \Box_{\bar{\gamma}} ] e^{2\xi\phi},
\end{split}\end{equation}

The canonical momenta in the dual frame can be obtained from the above lagrangian as
\begin{equation}
\bar{\pi}^{ij} = \frac{\delta L}{\delta \dot\bar\gamma_{ij}},\quad
\bar{\pi}^i_I = \frac{\delta L}{\delta \dot{A}_i^I},\quad
\bar\pi_\phi = \frac{\delta L}{\delta \dot\phi},\quad
\bar\pi_{\chi^a}=\frac{\delta L}{\delta \dot{\chi}^a},
\end{equation}
with conjugate momenta of the non-dynamical fields, $\bar{N}$, $\bar{N}_i$ and $A_r$ being zero. 

Expressing them in terms of quantities in the Einstein frame one gets
\begin{equation}\begin{split}
\bar\pi^{ij} &= \frac{1}{2\kappa^2} \sqrt{-\gamma} e^{2\xi\phi} (K\gamma^{ij} - K^{ij} ),\quad
\bar\pi^i_I = -\frac{2}{\kappa^2} \sqrt{-\gamma} \Sigma_{IJ} \gamma^{ij} F_{rj}^I,\\
\bar\pi_\phi &= \frac{1}{\kappa^2} \sqrt{-\gamma} (2 \xi K - \alpha \partial_r\phi),\quad
\bar\pi_{\chi^a} = - \frac{1}{\kappa^2} \sqrt{-\gamma} Z \partial_r\chi^a.
\end{split}\end{equation}

These expressions evaluated around the background in linearised order of perturbations in metric and other fields reduce to the following expressions.
\begin{equation}\begin{split}
\pi^{ta} = \frac{1}{4\kappa^2} e^{2\xi\phi_B} e^{-3A}f^{-1/2} \partial_r (e^{4A} S_t^a),\\
\pi^a_I = -\frac{2}{\kappa^2} e^A f^{1/2} \Sigma_{IJ} (\partial_r a_a^J + f^{-1} (\partial_r a_t^J) S_t^a ),\\
\pi_{\chi^a} = - \frac{1}{\kappa^2} e^{3A} f^{1/2} Z \partial_r\tau^a.
\end{split}\end{equation}

In order to make connection to the asymptotic expressions we will express the above equations in terms of $\Theta^a$, $a_a^I$ and $\tau^a$. We will consider only the expression in zeroeth order of $\omega$. Furthermore, we will use the radial coordinate $v$ instead of $r$. Substituting the background values of the fields and using $dr = -sgn(\theta) v^{-\theta/2}{\mathcal F}^{-1/2}(v) \frac{dv}{v}$ we obtain,
\begin{equation}\begin{split}
\pi^{ta} &= - \frac{sgn(\theta)}{4\kappa^2} v^{\theta-z-1}\partial_v ( v^{4-2\theta} (\Theta^{a(0)} + \frac{i\omega}{p} \tau^{a(0)})),\\
\pi_1^a &= \frac{sgn(\theta)}{2\kappa^2} [v^{z+\theta-3} F(v)\partial_va_a^{1(0)} + 4 sgn(\theta) q_1 (\Theta^{a(0)} + \frac{i\omega}{p} \tau^{a(0)})],\\
\pi_2^a &= \frac{sgn(\theta)}{2\kappa^2} [v^{3z+\theta-1} F(v)\partial_va_a^{2(0)} + 4 sgn(\theta) q_2 (\Theta^{a(0)} + \frac{i\omega}{p} \tau^{a(0)})],\\
\pi_{\chi^a} &= \frac{i\omega}{2p\kappa^2} [-sgn(\theta) v^{5-z-\theta}\partial_v\Theta^{a(0)} - 4 q_I a_a^{I(0)}].
\end{split}\end{equation}
Substituting the expressions for the fields in small frequency limit we can obtain the expressions of the canonical momenta. As has been explained in \cite{Cremonini:2018jrx} the asymptotic expressions provide a map between the two sets.  One set is given by the fluctuations, $\Theta^{a(0)}$, $a_a^{I(0)}$, $\tau^{a(0)}$ along with their conjugate momenta. The other set consists of the modes $\Theta_1^a$, $\Theta_2^a$, $a_{a0}^a$, $C_I^a$ and $\tau^a$. 

The set of fluctuations 
should be identified with the local sources and operators in the boundary theory but with these expressions they will not be independent of radial variable $v$. In order to identify the local sources and operators one needs to consider holographic renormalisation of the action. Since our case is very similar to \cite{Cremonini:2018jrx} we refer their analysis for details. This identification involves a canonical transformation among the fluctuations and their conjugate momenta, which can be realised by adding appropriate counterterms in the regularised action.  
The canonical transformation, in absence of magnetic field has been described elaborately In \cite{Cremonini:2018jrx}. They have considered on shell regularised action for the model with the black hole solution as the background. Through addition of counterterms at the boundary the variables $\pi^{ta}$, $A_1^a$ and $\pi^{\chi^a}$ undergo canonical transformations, keeping $A_2^a$ and its canonical conjugate momentum unchanged. 

As has been mentioned earlier, since the effect of the magnetic field appears at the linear order in frequency or higher, small frequency expansion of the fluctuations $\Theta^{a(0)}$, $a_a^{I(0)}$, $\tau^{a(0)}$ remain the same as in the case of zero magnetic field. However, there are differences in the expression of the blackening factor ${\mathcal F}(v)$ and so the counterterms will be modified in this case. In presence of magnetic field we are assuming one can make a similar canonical transformation through addition of counterterms and obtain the transformed variables which are appropriate to make identification of the local sources and operators on the boundary. A similar addition of counterterms will give rise to the following asymptotic expression of the transformed variables,
\begin{equation}\begin{split}\label{asymp-renorm-variable}
\Pi^{ta} = - \frac{1}{4\kappa^2} v^{-2z} (\Theta_2^a + 4 \mu^I C_I^a) +...,\quad\quad
{\mathfrak a}_a^1 = a_{a0}^1 - \mu^1 \Theta_1 + ...,\\
\Pi_{\chi^a} = \frac{-2i\omega}{p\kappa^2} q_I {\mathfrak a}_a^I + ...,\quad\quad
{\mathfrak a}_a^2 = a_{a0}^2 - \mu^2 \Theta_1 + ...,
\end{split}\end{equation}
where the chemical potentials are given by
\begin{equation}\label{mu-value}
\mu^1 = -\frac{4 sgn(\theta) q_1 v_h^{2+z-\theta}}{2+z-\theta},
\quad
\mu^2 = -\frac{4 sgn(\theta) q_2 v_h^{\theta - z}}{\theta - z}.
\end{equation}
These transformed variables are related to the original symplectic variables through a canonical transformation. Following \cite{Cremonini:2018jrx} we identify the asymptotic expressions of these transformed variables with the observables in the dual field theory as follows. One can define different holographically dual theory by imposing different boundary conditions. For Dirichlet boundary condition on $A_1^a$, which requires addition of an additional boundary term to the on shell action along with counterterms \cite{Cremonini:2018jrx}, the observables and the sources for energy flux are given by
\begin{equation}\begin{split} \label{Ea}
{\mathcal E}^a = 2 \lim\limits_{\bar{r}\rightarrow\infty} e^{2z\bar{r}}\Pi^{ta} = -\frac{1}{2\kappa^2} (\Theta_2^a + 4 \mu^I C_{I0}^a ),\quad\quad
\Theta^a_1 = \lim\limits_{\bar{r}\rightarrow\infty} e^{-2z\bar{r}} n_a,
\end{split}\end{equation}
respectively where $\bar{r}$ is related to $r$ through $r\sim \frac{2}{|\theta|} e^-\frac{\theta \bar{r}}{2}$ and   $n_a$ is the shift function in the decomposition of the metric $\bar{\gamma}_{ij}$ as $\bar{\gamma}_{ij} dx^i dx^j = -(n^2 - n_an^a)dt^2 + 2n_a dt dx^a + \sigma_{ab}dx^adx^b$, $a,b=1,2$. 
Similarly the observable for $U(1)$ currents and pseudoscalars are given by
\begin{equation}\begin{split} \label{JIa}
{\mathcal J}_I^a = \lim\limits_{\bar{r}\rightarrow\infty} \Pi^a_I = -\frac{2}{\kappa^2} (C_{I0}^a - \frac{i\omega q_I}{p} \tau_0^a ),\quad\quad
{\mathcal X}_a = \lim\limits_{\bar{r}\rightarrow\infty} \Pi_{\chi^a} = - \frac{2i\omega}{p\kappa^2} q_I \alpha_a^I,
\end{split}\end{equation}
respectively and $\alpha_a^I$ is given by 
\begin{equation}\label{alpha}
\alpha_a^I = a_0^I - \mu^I\Theta_1^a.
\end{equation}

\end{document}